\documentclass[twoside,twocolumn,english,aps,manuscript,aps,preprint]{revtex4-1}
\usepackage[T1]{fontenc}
\usepackage[latin9]{inputenc}
\usepackage{fancyhdr}
\usepackage{graphicx}
\usepackage{esint}

\makeatletter

\@ifundefined{textcolor}{}
{%
 \definecolor{BLACK}{gray}{0}
 \definecolor{WHITE}{gray}{1}
 \definecolor{RED}{rgb}{1,0,0}
 \definecolor{GREEN}{rgb}{0,1,0}
 \definecolor{BLUE}{rgb}{0,0,1}
 \definecolor{CYAN}{cmyk}{1,0,0,0}
 \definecolor{MAGENTA}{cmyk}{0,1,0,0}
 \definecolor{YELLOW}{cmyk}{0,0,1,0}
}

\makeatother

\usepackage{babel}
\begin{document}


\title{Characterization of a metastable neon beam extracted from a commercial
RF ion source}

\author{B. Ohayon}

\affiliation{Racah Institute of Physics, Hebrew University of Jerusalem, Jerusalem
91904, Israel}

\author{E. Wåhlin}

\affiliation{Beam Imaging Solutions, 1610 Pace St., Longmont 80504, USA}

\author{G. Ron}

\email[Author to whom correspondence should be addressed. Electronic mail: ]{gron@racah.phys.huji.ac.il }

\affiliation{Racah Institute of Physics, Hebrew University of Jerusalem, Jerusalem
91904, Israel}

\date{\today}
\begin{abstract}
We have used a commercial RF ion-source to extract a beam of metastable
neon atoms. The source was easily incorporated into our existing system
and was operative within a day of installation. The metastable velocity
distribution, flux, flow, and efficiency were investigated for different
RF powers and pressures, and an optimum was found at a flux density
of $2\times10^{12}\,$atoms/s/sr. To obtain an accurate measurement
of the amount of metastable atoms leaving the source, we insert a
Faraday cup in the beam line and quench some of them using a weak
$633\,$nm laser beam. In order to determine how much of the beam
was quenched before reaching our detector, we devised a simple model
for the quenching transition and investigated it for different laser
powers. This detection method can be easily adapted to other noble
gas atoms.
\end{abstract}
\maketitle

\section{introduction}

In 2001, two groups realized the first Bose-Einstein condensate of
metastable helium \cite{2001-Aspect-HeliumBEC,2001-HeBEC}. To excite
it to the metastable state, they have both used a high voltage discharge
source \cite{2000-Aspect-DCsource,2001-Leduc-DC-source}, where atoms
are excited to upper states, by running a DC discharge through the
expanding atomic beam, which later decayed to a long-lived metastable
state. In recent years, there has been much interest in experiments
with ultracold metastable noble gasses (See \cite{2012-Cold-Metastable-review}
for a review). Also, advances in the efficiency of production and
trapping of metastable noble gasses have enabled spectroscopy and
abundance measurements with tiny fractions of rare isotopes. The isotope
shift and hyperfine structure of isotopes of argon \cite{2008-Argon-Blaum-Isotope-Shift},
krypton \cite{1990-Krypton-Isotope-Shofts}, neon \cite{2011-isotope-shift}
and xenon \cite{1993-Isoshift-Xenon} were measured. Moreover, isotope
shift measurements have determined the $\mbox{\ensuremath{^{3}}He}$
nuclear charge radius \cite{1991-he34-IsoShift} and recently those
of $\mbox{\ensuremath{^{6,8}}He}$ \cite{2013-laserProbing-nuclei-Lu}.
In atomic traps, a novel technique for radio-dating using abundance
measurements of noble gas isotopes called Atom Trap Trace Analysis
(ATTA) has been introduced with krypton \cite{1999-ATTA-Krypton},
and recently used with argon \cite{2013-ArgonATTA}. Lastly, there
is also a growing interest in measurement of cold and trapped short-lived
isotopes of the noble gasses neon and helium \cite{2014-Behr-beta}.
\begin{figure*}
a)\includegraphics[scale=0.2]{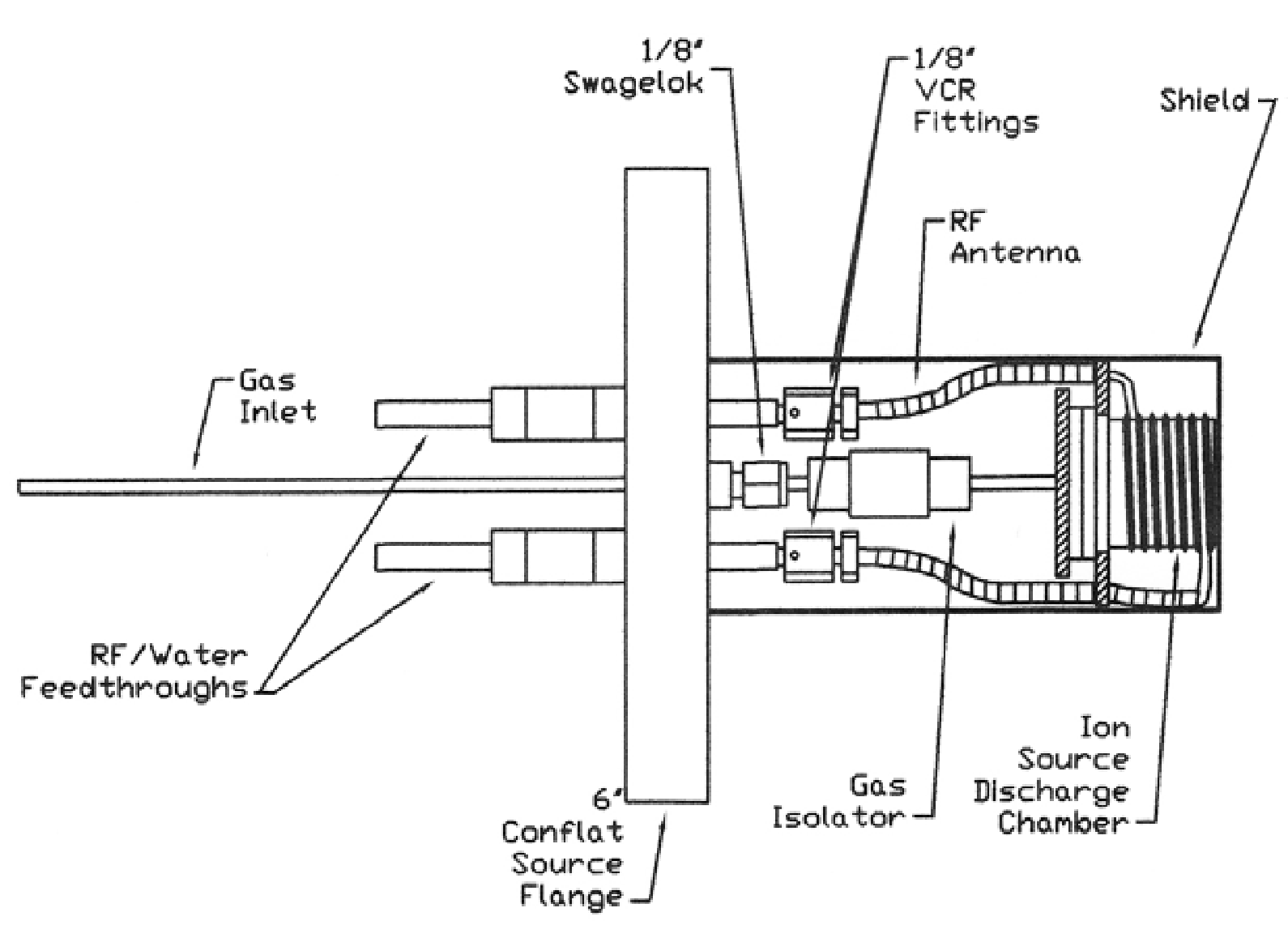}b)\includegraphics[scale=0.34]{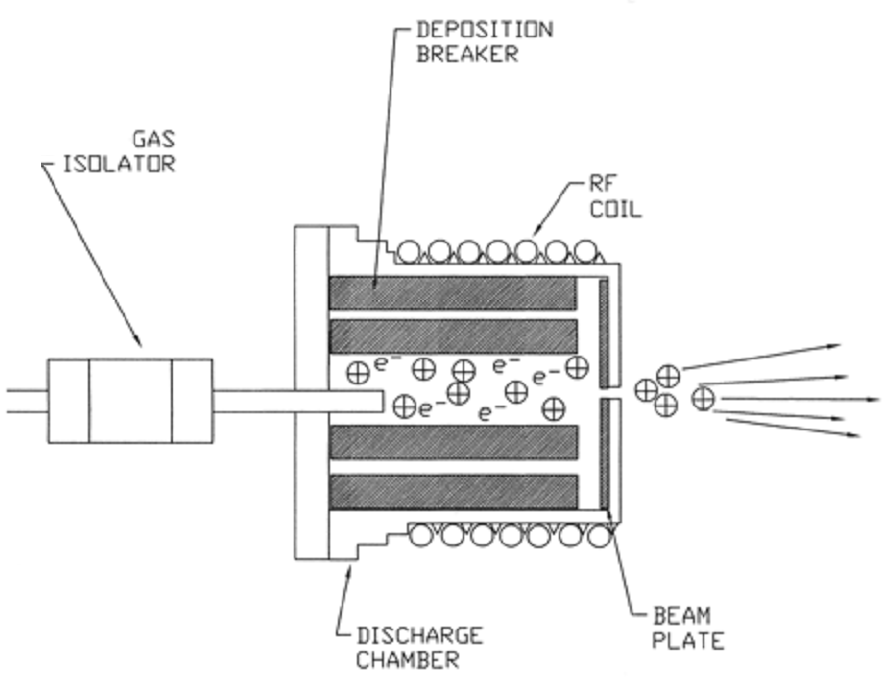}

\protect\caption{a) RFIS-100 ion source schematics. b) Close up on discharge chamber.
Both are adapted from \cite{2011-RFIS-Guide}.}

\label{Fig: Schematics}
\end{figure*}

These developments have sparked interest in developing efficient metastable
sources which are simple to construct, implement and maintain. Exotic
sources such as an inverted magnetron pressure gauge \cite{Magnetron-2004},
and an all optical source \cite{2011-VUV-Krypton} have been constructed,
but the simplest and most efficient sources today usually rely on
an RF discharge \cite{2001-RF-Discharge-Krypton-Chen}. Despite needing
much less maintenance than DC discharge sources, home built RF discharge
sources degrade over time due to depositions on the discharge tube
from sputtering of stainless steel vacuum parts. After experiencing
difficulties in continuously operating a home built RF source we decided
to acquire a commercial ion source (BIS RFIS-100) and investigate
the conditions for efficiently extracting metastable neon atoms from
it. In this paper we review our findings.

\section{RFIS-100 ion source}

In Fig. \ref{Fig: Schematics}a, adapted from \cite{2011-RFIS-Guide},
we show an overview of the source design. neon flows through the gas
inlet, is delayed in the ceramic gas isolator to prevent the plasma
short-circuiting to ground, and is excited in the alumina discharge
chamber (Fig. \ref{Fig: Schematics}b), where a discharge is created
by applying high RF power at $13.56\mbox{ MHz}$ to a helical resonator
copper antenna. The discharge chamber is equipped with an alumina
deposition breaker, which serves as to prevent ablated metal particles
from forming a conductive coating on the chamber which could shield
the discharge from the RF power. A molybdenum beam plate acts as a
nozzle which can be easily drilled to different diameters. The source
also incorporates a piezoelectric high voltage module to help with
ignition of the discharge.

There are normally two modes of operation with RF sources. Namely,
capacitively coupled plasma, where electrons accelerated by the RF
electric field ionize the gas to cause an avalanche, and inductively
coupled plasma (ICP), where the electrons are accelerated by time
varying magnetic fields \cite{2000ICP-CCP}. ICP is initiated at high
powers and pressures and is considered desirable in terms of metastable
production \cite{2009-Florian}. The signature of the ICP discharge
is the requirement to re-tune the RF antenna's impedance after the
discharge starts. Good impedance matching between the antenna network
and the RF power supply (T\&C AG 0613 600W) is achieved via two high
voltage vacuum variable capacitors that are connected in parallel
and in series with the antenna. At virtually all pressures and powers
in which neon ignites we observed a drastic change in impedance which
indicates the formation of an ICP.

Igniting the plasma is harder than maintaining it, for this reason
we usually start at high pressure and high RF power (\textbf{$\sim300\,$}W)
and then quickly ramp down the power. Also, it is harder to start
the plasma when the source is hot. Even though the hollow antenna
is cooled by pumping a high pressure refrigeration fluid (tetrafluoromethane),
the $13.56\,\mbox{MHz}$ radiation dissipates significant power in
the vacuum parts, and above $250\mbox{\,\ W}$, care must be taken.
Thus, the power available to the discharge is probably less than the
forward power measured by the generator (also reported at \cite{2011-Welte}).
At high pressures of a few $10^{-5}\,$Torr, pure neon lights up easily
and plasma can be sustained at powers as low as about $150\mbox{\,\ W}$.
At low pressures of a few $10^{-6}\,$Torr, plasma can only be sustained
at high powers of about $220\,$W. When mixing a small amount of xenon
(Similar to reports by \cite{2011-Suil}), the plasma ignites at virtually
all powers and pressures.

\section{Light induced quenching of metastable neon}

The steady-state scattering rate of classical radiation by a two level
system with natural linewidth $\Gamma$ is \cite{1985-Metcalf-Book}

\begin{equation}
r\left(\vec{v}\right)=\frac{s_{0}/2}{1+s_{0}+(2\delta\left(\vec{v}\right)/\Gamma)^{2}},\label{eq:force}
\end{equation}
where $s_{0}=I_{l}/I_{s}$ is the saturation parameter, $I_{l}$ is
the laser intensity, and $I_{s}$ the saturation intensity. The detuning
is given by
\begin{equation}
\delta\left(\vec{v}\right)=\delta_{0}-\vec{k}\cdot\vec{v}\label{eq: Doppler}
\end{equation}
where $\delta_{0}$ is the difference between the laser and the atomic
transition frequency, and $\vec{k}\cdot\vec{v}$ the Doppler detuning,
caused by the relative velocity of the light and atom.

\begin{figure}
\includegraphics[scale=0.4]{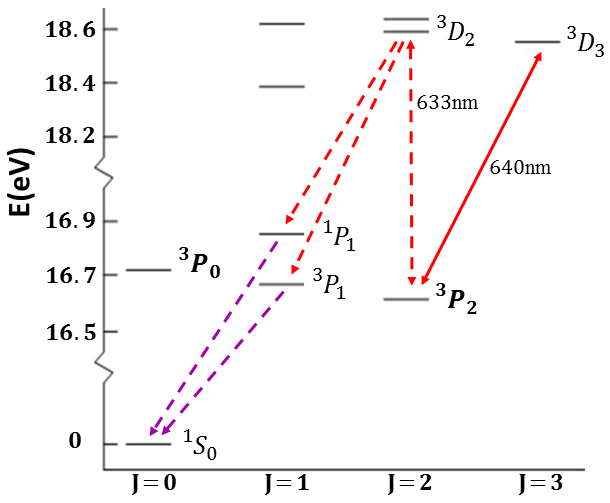}

\protect\caption{Relevant transition lines of neon \cite{2014-NIST_ASD}. The metastable
levels $^{3}P_{0,2}$ are indicated in bold. Upward facing arrows
represent transitions coupled by our lasers, with the wavelength indicated.
Downward facing arrows represent emission. The solid line represents
the cycling transition and the broken lines represent the quenching
route.}

\label{Fig: Levels}
\end{figure}

Noble gasses possess a long lived metastable state which forms such
a two-level system \cite{2012-Cold-Metastable-review}. With neon,
the cycling transition is between the levels $^{3}P_{2}-{}^{3}D_{3}$,
as shown in Fig. \ref{Fig: Levels}. Its wavelength in air is $640.2\,\mbox{nm}$.
It has a linewidth of $\Gamma=8.2\left(2\pi\right)\,\mbox{MHz}$,
or, using eq. \ref{eq: Doppler}, $\sigma_{\Gamma}=\Gamma/k=5.2\mbox{\,\ m/s}$
in units of velocity. Another transition of interest is $^{3}P_{2}-{}^{3}D_{2}$
which has a wavelength of $633.4\,\mbox{nm}$ \cite{2008-NIST}. It
is not a closed transition, since the $^{3}D_{2}$ state can decay
via a dipole transition to the $^{3}P_{1}$ and $^{1}P_{1}$ states
(which decay immediately to the ground state) as well as to $^{3}P_{2}$.
Thus a laser tuned to this wavelength can deplete the population in
the $^{3}P_{2}$ state; this process is also called \textit{quenching}.
The states of interest, and the relevant wavelengths are presented
in Fig. \ref{Fig: Levels}.

When a metastable beam encounters a laser beam tuned to the quenching
transition, the atoms have a chance of absorbing photons and deexciting
from the metastable state. The longer an atom spends inside the quench
beam, and the stronger the laser, the chances of its survival decrease
exponentially. For a transverse laser locked to the quenching transition,
the detuning from the atomic transition is governed by the atom's
transverse velocity (eq. \ref{eq: Doppler}). We thus model, using
eq. \ref{eq:force}, this average survival rate of a metastable atom,
with $v_{r}$ and $v_{z}$ the transverse and longitudinal velocities,
which travels through a weak transverse quenching beam can be written
as 

\begin{equation}
f_{Q}\left(v_{z},v_{r}\right)=exp\left(-\frac{v_{q}}{v_{z}}\frac{1}{1+(2v_{r}/\sigma_{q})^{2}}\right).\label{eq: quench_dist}
\end{equation}

\begin{figure*}
\includegraphics[scale=0.4]{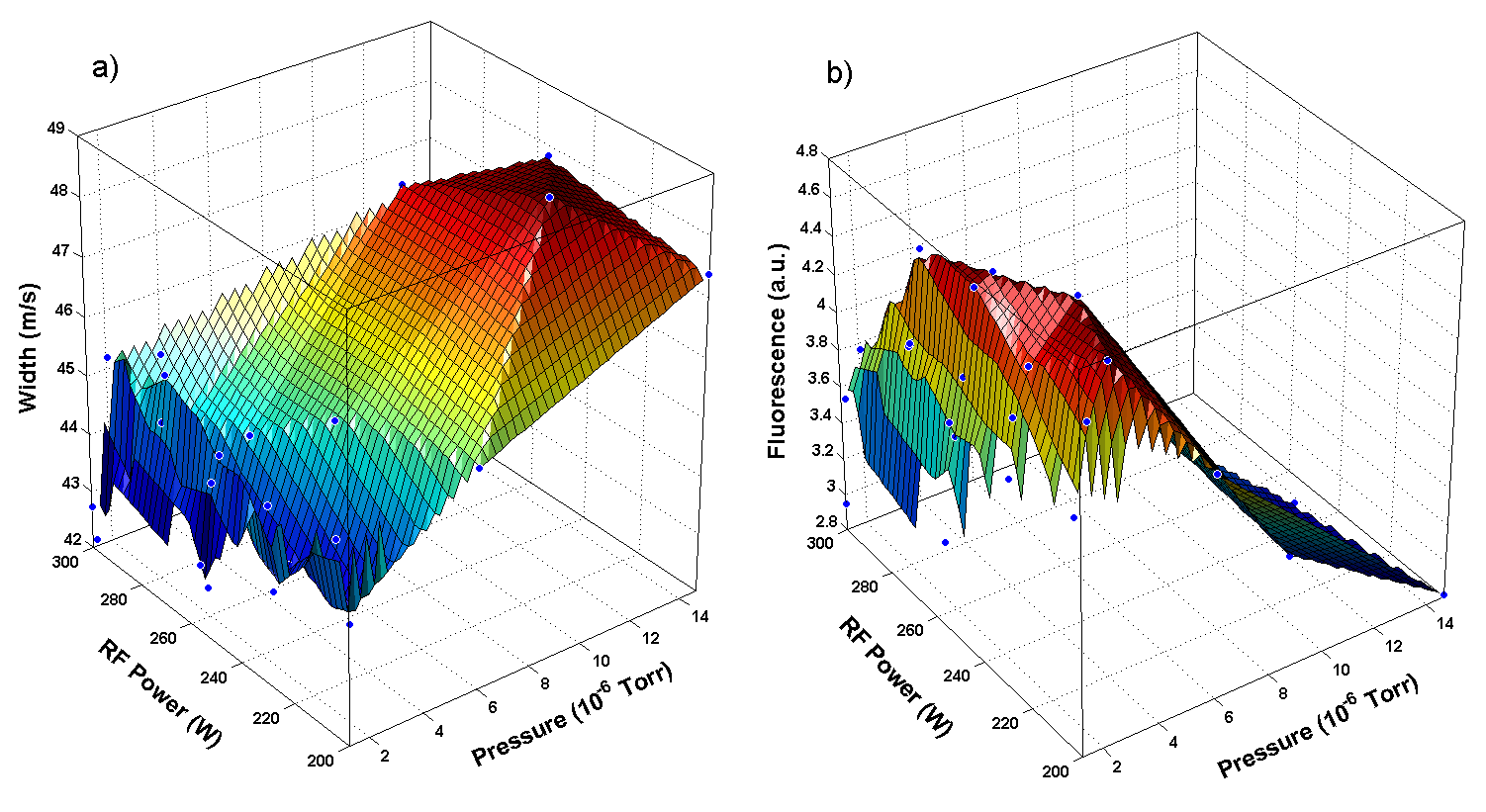}

\protect\caption{a) Transverse velocity width as a function RF power and chamber pressure.
b) Total fluorescence as a function RF power and chamber pressure.}

\label{Fig: Transverse}
\end{figure*}
Where $v_{q}$ is the effective intensity of the quench laser, and
$\sigma_{q}$ is an effective transition linewidth, both in units
of velocity. Since our source incorporates a discharge chamber with
a small aperture (Fig. \ref{Fig: Schematics}), the emerging beam
is assumed to have an effusive velocity distribution \cite{pauly2000atom}:\\
\begin{eqnarray}
f_{B}\left(v_{r},v_{z}\right) & = & \frac{9}{2}\frac{v_{z}^{3}}{v_{mp}^{4}}\mbox{exp}\left(-\frac{3}{2}\frac{v_{z}^{2}}{v_{mp}^{2}}\right)\times\nonumber \\
 &  & \frac{1}{\sigma_{r}^{2}}\mbox{exp}\left(-\frac{v_{r}^{2}}{2\sigma_{r}^{2}}\right),
\end{eqnarray}
which is cylindrically normalized such that $\int_{0}^{\infty}\int_{0}^{\infty}f_{B}\left(v_{r},v_{z}\right)v_{r}\mathrm{d}v_{r}\mathrm{d}v_{z}=1$.
$v_{mp}$ is the most probable velocity, and $\sigma_{r}$ the transverse
rms velocity. The total velocity distribution of an atomic beam which
passed a transverse laser tuned to the quenching transition is thus
\[
f_{tot}\left(v_{r},v_{z}\right)=f_{B}\left(v_{r},v_{z}\right)f_{Q}\left(v_{z},v_{r}\right).
\]
Adding a weak probe beam tuned to the cycling transition and collecting
the fluorescence using a photomultiplier tube (PMT), we measure a
signal proportional to 

\begin{equation}
F\left(v_{r}\right)=\int_{0}^{\infty}f_{tot}\left(v_{z},v_{r}\right)\mathrm{d}v_{z}\label{eq: Scope-Signal trans}
\end{equation}
for a transverse probe beam, and to
\begin{equation}
F\left(v_{z}\right)=\int_{0}^{\infty}f_{tot}\left(v_{z},v_{r}\right)v_{r}\mathrm{d}v_{r}\label{eq:Scope_signal: long}
\end{equation}
for a longitudinal probe.

To investigate the working conditions of the source, we use a weak
transverse beam which probes the cycling transition and record the
PMT signal for different RF powers and neon pressures. The scan voltage
to frequency calibration was accomplished by conducting a wide scan
and observing the isotope shift on a saturated absorption setup \cite{2013-Boaz}.
Since the fluorescence is visible to the naked eye, we noticed a slight
fluorescence inside the entire source chamber which is unrelated to
the atomic beam. Since we work with high RF power, we conclude that
some of the background gas is also excited. This means that our PMT
signal is comprised of a narrow atomic beam signal atop a wider background
signal. This assumption is validated when we use the quenching beam.

\begin{figure}
\includegraphics[scale=0.37]{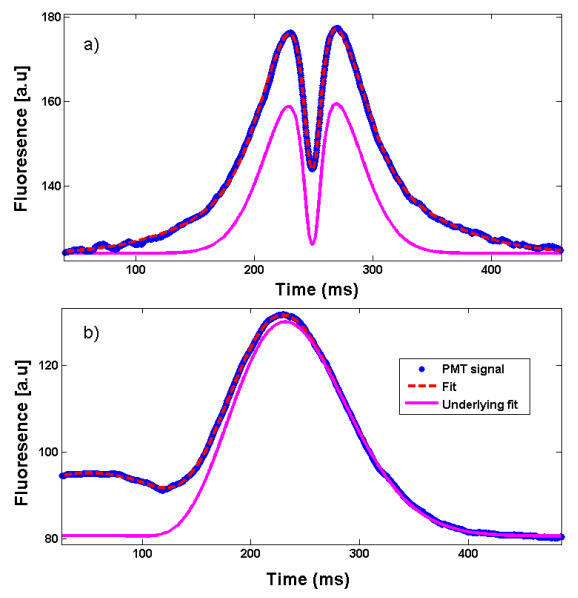}

\protect\caption{Example of fluorescence quenching fit at $3\,\mbox{mW}$ laser power.
a) Transverse probe scope signal, fitted to eq. \ref{eq: Scope-Signal trans}
(dashed line). b) Longitudinal probe signal, fitted to eq. \ref{eq:Scope_signal: long}.
Solid lines are the fits with the background Gaussian subtracted.}

\label{fig: typ-Quench Fluor}
\end{figure}
In Fig \ref{Fig: Transverse}, we show the results obtained from fitting
the PMT signal to a double Gaussian for different working conditions.
Even though the atomic flux increases with pressure, high chamber
pressure tend to quench the metastable beam through collisional deexcitations
inside as well as outside of the source \cite{2010-Welte-florian-Paper}.
The optimum flux is at a chamber pressure of about $5\times10^{-6}\,\mbox{Torr}$
and does not change much with RF power, which has to be above $200\,\mbox{W}$
to sustain plasma at such low pressures. Also, a small increase of
the transverse velocity is observed at high pressures, the transverse
rms velocity at the desired pressure is $\sigma_{r}=45\,\mbox{m/s}$.

\begin{figure}
\includegraphics[scale=0.17]{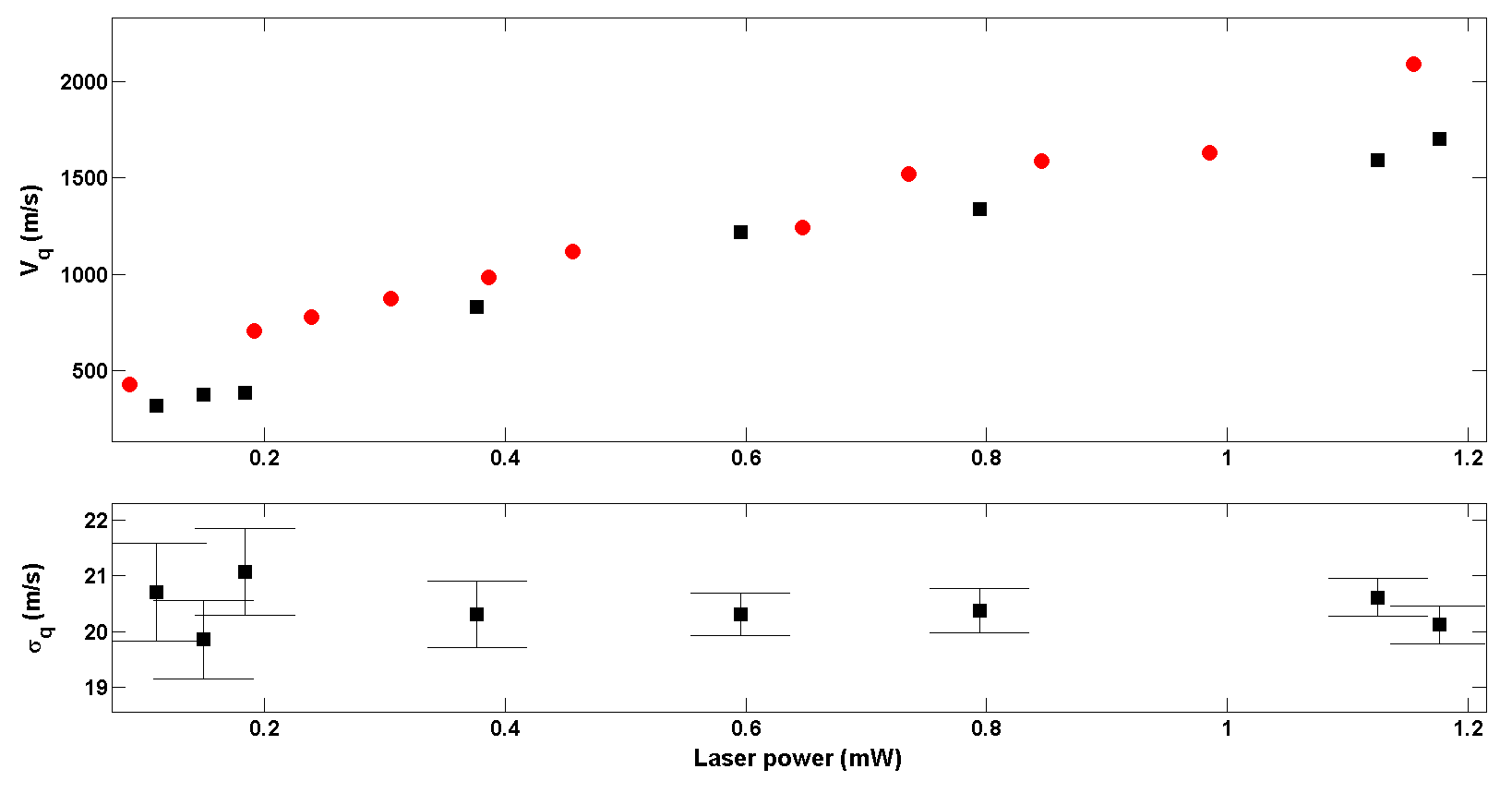}

\protect\caption{(Top) Effective quench laser power $v_{q}$ for transverse (squares)
and longitudinal (circles) probing. Markers are larger than the errorbars.
(Bottom) Effective transition linewidth $\sigma_{q}$ for transverse
probing.}

\label{Fig: quench parameters}
\end{figure}

To investigate the properties of the quenching transition, we add
a large transverse $633\,$nm beam immediately after the source aperture,
which quenches some of the metastable atoms before they reach the
light collection region. For a transverse probe beam, we observe that
the quenching beam affects the metastable atomic beam signal without
affecting the background signal. We thus fit to a wide Gaussian plus
transverse quenching (eq. \ref{eq: Scope-Signal trans}). A typical
fit, as well as the underlying fit with the background subtracted,
is shown in Fig. \ref{fig: typ-Quench Fluor}a. The same was done
with a longitudinal probe beam and fit to eq. \ref{eq:Scope_signal: long}
and is shown in Fig. \ref{fig: typ-Quench Fluor}b. We recorded and
fit the transverse and longitudinal quenched fluorescence signal for
different quench laser powers. The results are shown in Fig. \ref{Fig: quench parameters}.
As modeled in eq. \ref{eq: quench_dist}, the effective quench laser
power $v_{q}$ at low quench powers is linear with laser power and
gives similar results for both the transverse and longitudinal probes.
Also, the effective transition linewidth $\sigma_{q}$ is constant
over a wide range of quench laser power.

\section{Metastable detection using a Faraday-cup}

Schemes for detecting metastable atoms usually rely on their long
lifetime and large stored energy \cite{1996-hotop}. When metastable
atoms collide with most surfaces, they immediately ionize them. If
the surface is a conductor, a measurement of the ionization current
can be used to determine the flux of metastable atoms in a beam. 

In neon, the metastable states have approximately $E^{*}=16\,\mbox{eV}$
of internal energy, and the ionization energy is $E^{+}\sim22\,\mbox{eV}$.
When impacting on a metal surface with work function $\Phi$, which
for stainless steel is under $4.7\,\mbox{eV}$ \cite{1996-hotop},
two energetic conditions are met: $E^{*}>\Phi$ and $E^{+}\geq2\Phi$,
which enable two ionization processes by an exchange of electrons
between the metal and the atom \cite{1944-Metastable-Metal}. The
absolute metastable flux $F$ can be determined through \cite{1975-dunning}
\begin{equation}
I=e\gamma F\Delta Y,\label{eq:flux}
\end{equation}

where $I$ is the ionization current, $e$ is the fundamental charge,
$\gamma=0.61$ the emission coefficient for stainless steel impacted
by Ne{*} \cite{1975-dunning}, and $\Delta Y$ is the fraction of
atoms detected.

\begin{figure*}
\includegraphics[scale=0.52]{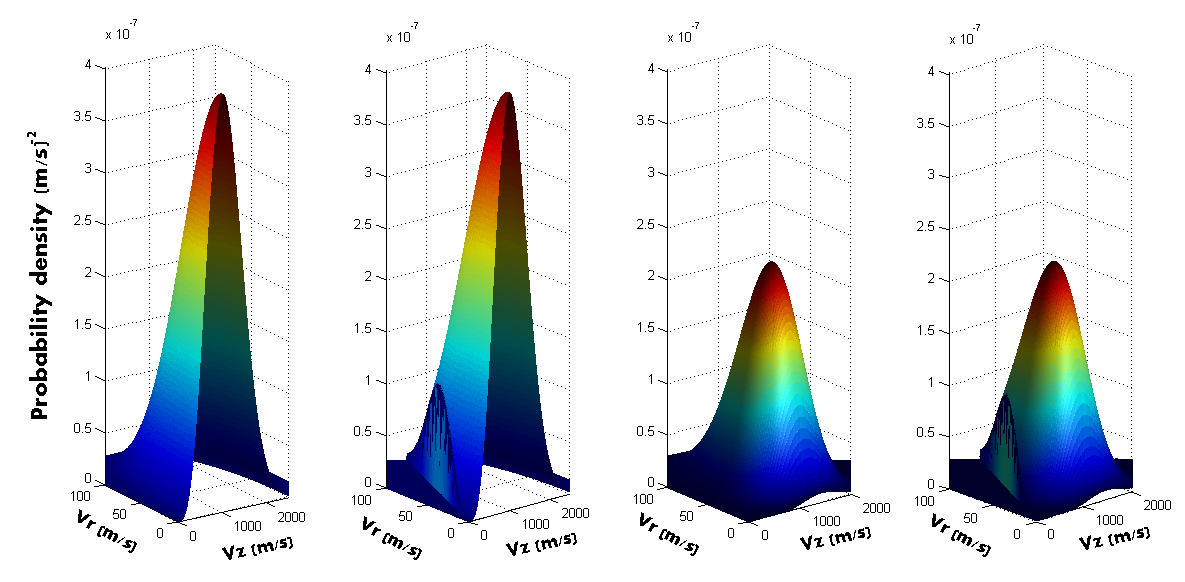}

\protect\caption{Probability density of transverse and longitudonal velocity distributions.
Left to right: Distribution of metastables leaving the source $\left(100\%\right)$,
Metastables that reach the Faraday-cup $\left(84\%\right)$, Metastables
left after crossing the quench laser beam $\left(79\%\right)$, Metastables
that reach the cup after crossing quench beam $\left(64\%\right)$}

\label{fig: Probaibilty dist velocities}
\end{figure*}
To implement this detection scheme we use a design similar to that
of \cite{2005-Ashmore}, for which a stainless steel detector plate
is mounted, along with an electron collecting cathode, along the beam
line. Upon operating the source, a large and noisy negative current
$\left(5\,\mu\mbox{A}\right)$ was measured. To stabilize the current,
a transverse magnetic field $\left(50\,\mbox{Gauss}\right)$ was introduced
by mounting a coil on the vacuum window of the source chamber and
flowing a large current $\left(\sim10\,\mbox{A}\right)$. The field
deflects fast electrons and ions before they reach the cup. To establish
how much of the current left results from atoms in the $^{3}P_{2}$
state only, we measure it again with the population depleted by quenching
it using the quench beam and subtract the results. This method is
similar to that used by \cite{1992-Missoury-Sat-Quench} for neon,
but since we did not need to quench all of the atomic beam, we could
use a quench laser two orders of magnitude weaker.

Based on the transverse and longitudinal velocity distributions, and
the distance and radius of the cup, we calculate the fraction of metastables
that reach the cup and arrive at the fraction $Y_{0}=0.84$. We now
simulate the amount of metastables reaching the cup when the quenching
laser is on and at maximal power using the parameters from Fig. \ref{fig: typ-Quench Fluor},
and arrive at the fraction $Y_{1}=0.64$. The fraction of metastables
which reach the cup and are quenched by the laser is thus $\Delta Y=Y_{0}-Y_{1}=0.20$.
The maximal difference current obtained using the Faraday cup was
$I=8\,\mbox{nA}$, using eq. \ref{eq:flux} we calculate that the
metastable flow is $4.2\times10^{11}\,$Ne{*}/s. The average flux
density in forward direction is then $2.3\times10^{12}\,$Ne{*}/s/sr.
The flow and flux depend strongly on the pumping speed, our source
was pumped by two turbo-molecular pumps which has a combined pumping
speed of $400\,$l/s. The total flow was measured using a flow meter
to be $1.4\,$sccm or $6.5\times10^{17}\,$Ne/s. Which yields an efficiency
of about $10^{-6}$. This is the efficiency at maximum metastable
flux. By operating the source with xenon, and mixing small amounts
of neon $\left(<10^{-6}\,\mbox{Torr}\right)$, a factor of 2-3 is
obtained with the efficiency and $10-20
$ of the flux is lost.

\section{Summary and outlook}

We have devised a simple, \textit{ad hoc} model to investigate the
amount of metastable neon atoms which are left after crossing a laser
beam tuned to a quenching transition. This model was used to investigate
how much of the ionization current in a Faraday-cup resulted from
atoms in a specific metastable state and so determine the metastable
flux density. Combined with spectroscopic measurements of the velocity
distribution, and a measurement of the atomic flow, a complete picture
of our source was obtained. Since the energy levels of other noble
gasses are similar, this detection method can be readily used in existing
metastable systems. The simplicity of incorporating, operating, and
maintaining a commercial source might make it a choice for future
industrial applications such as metastable atom lithography \cite{2010-helium-lithography}.

\section{Acknowledgments}

The authors thank M. Oberthaler and the Synthetic Quantum Systems
Group at the Kirchhoff Institute for Physics at the University of
Heidelberg. This work was supported by the Israeli Science Foundation
under ISF grant 177/11. Ben Ohayon is supported by the Hoffman Leadership
and Responsibility fund.

\bibliographystyle{apsrev}
\bibliography{ALL}

\begin{thebibliography}{34}
\expandafter\ifx\csname natexlab\endcsname\relax\def\natexlab#1{#1}\fi
\expandafter\ifx\csname bibnamefont\endcsname\relax
  \def\bibnamefont#1{#1}\fi
\expandafter\ifx\csname bibfnamefont\endcsname\relax
  \def\bibfnamefont#1{#1}\fi
\expandafter\ifx\csname citenamefont\endcsname\relax
  \def\citenamefont#1{#1}\fi
\expandafter\ifx\csname url\endcsname\relax
  \def\url#1{\texttt{#1}}\fi
\expandafter\ifx\csname urlprefix\endcsname\relax\def\urlprefix{URL }\fi
\providecommand{\bibinfo}[2]{#2}
\providecommand{\eprint}[2][]{\url{#2}}

\bibitem[{\citenamefont{Robert et~al.}(2001)\citenamefont{Robert, Sirjean,
  Browaeys, Poupard, Nowak, Boiron, Westbrook, and
  Aspect}}]{2001-Aspect-HeliumBEC}
\bibinfo{author}{\bibfnamefont{A.}~\bibnamefont{Robert}},
  \bibinfo{author}{\bibfnamefont{O.}~\bibnamefont{Sirjean}},
  \bibinfo{author}{\bibfnamefont{A.}~\bibnamefont{Browaeys}},
  \bibinfo{author}{\bibfnamefont{J.}~\bibnamefont{Poupard}},
  \bibinfo{author}{\bibfnamefont{S.}~\bibnamefont{Nowak}},
  \bibinfo{author}{\bibfnamefont{D.}~\bibnamefont{Boiron}},
  \bibinfo{author}{\bibfnamefont{C.~I.} \bibnamefont{Westbrook}},
  \bibnamefont{and} \bibinfo{author}{\bibfnamefont{A.}~\bibnamefont{Aspect}},
  \bibinfo{journal}{Science} \textbf{\bibinfo{volume}{292}},
  \bibinfo{pages}{461} (\bibinfo{year}{2001}).

\bibitem[{\citenamefont{Pereira Dos~Santos et~al.}(2001)\citenamefont{Pereira
  Dos~Santos, L\'eonard, Wang, Barrelet, Perales, Rasel, Unnikrishnan, Leduc,
  and Cohen-Tannoudji}}]{2001-HeBEC}
\bibinfo{author}{\bibfnamefont{F.}~\bibnamefont{Pereira Dos~Santos}},
  \bibinfo{author}{\bibfnamefont{J.}~\bibnamefont{L\'eonard}},
  \bibinfo{author}{\bibfnamefont{J.}~\bibnamefont{Wang}},
  \bibinfo{author}{\bibfnamefont{C.~J.} \bibnamefont{Barrelet}},
  \bibinfo{author}{\bibfnamefont{F.}~\bibnamefont{Perales}},
  \bibinfo{author}{\bibfnamefont{E.}~\bibnamefont{Rasel}},
  \bibinfo{author}{\bibfnamefont{C.~S.} \bibnamefont{Unnikrishnan}},
  \bibinfo{author}{\bibfnamefont{M.}~\bibnamefont{Leduc}}, \bibnamefont{and}
  \bibinfo{author}{\bibfnamefont{C.}~\bibnamefont{Cohen-Tannoudji}},
  \bibinfo{journal}{Phys. Rev. Lett.} \textbf{\bibinfo{volume}{86}},
  \bibinfo{pages}{3459} (\bibinfo{year}{2001}).

\bibitem[{\citenamefont{Browaeys et~al.}(2000)\citenamefont{Browaeys, Poupard,
  Robert, Nowak, Rooijakkers, Arimondo, Marcassa, Boiron, Westbrook, and
  Aspect}}]{2000-Aspect-DCsource}
\bibinfo{author}{\bibfnamefont{A.}~\bibnamefont{Browaeys}},
  \bibinfo{author}{\bibfnamefont{J.}~\bibnamefont{Poupard}},
  \bibinfo{author}{\bibfnamefont{A.}~\bibnamefont{Robert}},
  \bibinfo{author}{\bibfnamefont{S.}~\bibnamefont{Nowak}},
  \bibinfo{author}{\bibfnamefont{W.}~\bibnamefont{Rooijakkers}},
  \bibinfo{author}{\bibfnamefont{E.}~\bibnamefont{Arimondo}},
  \bibinfo{author}{\bibfnamefont{L.}~\bibnamefont{Marcassa}},
  \bibinfo{author}{\bibfnamefont{D.}~\bibnamefont{Boiron}},
  \bibinfo{author}{\bibfnamefont{C.}~\bibnamefont{Westbrook}},
  \bibnamefont{and} \bibinfo{author}{\bibfnamefont{A.}~\bibnamefont{Aspect}},
  \bibinfo{journal}{The European Physical Journal D}
  \textbf{\bibinfo{volume}{8}}, \bibinfo{pages}{199} (\bibinfo{year}{2000}),
  ISSN \bibinfo{issn}{1434-6060}.

\bibitem[{\citenamefont{F.~Pereira Dos~Santos
  et~al.}(2001)\citenamefont{F.~Pereira Dos~Santos, Perales, Léonard, Sinatra,
  Wang, Pavone, Rasel, Unnikrishnan, and Leduc}}]{2001-Leduc-DC-source}
\bibinfo{author}{\bibfnamefont{n.}~\bibnamefont{F.~Pereira Dos~Santos}},
  \bibinfo{author}{\bibfnamefont{F.}~\bibnamefont{Perales}},
  \bibinfo{author}{\bibfnamefont{J.}~\bibnamefont{Léonard}},
  \bibinfo{author}{\bibfnamefont{A.}~\bibnamefont{Sinatra}},
  \bibinfo{author}{\bibfnamefont{J.}~\bibnamefont{Wang}},
  \bibinfo{author}{\bibfnamefont{F.~S.} \bibnamefont{Pavone}},
  \bibinfo{author}{\bibfnamefont{E.}~\bibnamefont{Rasel}},
  \bibinfo{author}{\bibfnamefont{C.~S.} \bibnamefont{Unnikrishnan}},
  \bibnamefont{and} \bibinfo{author}{\bibfnamefont{M.}~\bibnamefont{Leduc}},
  \bibinfo{journal}{The European Physical Journal Applied Physics}
  \textbf{\bibinfo{volume}{14}}, \bibinfo{pages}{69} (\bibinfo{year}{2001}),
  ISSN \bibinfo{issn}{1286-0050}.

\bibitem[{\citenamefont{Vassen et~al.}(2012)\citenamefont{Vassen,
  Cohen-Tannoudji, Leduc, Boiron, Westbrook, Truscott, Baldwin, Birkl, Cancio,
  and Trippenbach}}]{2012-Cold-Metastable-review}
\bibinfo{author}{\bibfnamefont{W.}~\bibnamefont{Vassen}},
  \bibinfo{author}{\bibfnamefont{C.}~\bibnamefont{Cohen-Tannoudji}},
  \bibinfo{author}{\bibfnamefont{M.}~\bibnamefont{Leduc}},
  \bibinfo{author}{\bibfnamefont{D.}~\bibnamefont{Boiron}},
  \bibinfo{author}{\bibfnamefont{C.~I.} \bibnamefont{Westbrook}},
  \bibinfo{author}{\bibfnamefont{A.}~\bibnamefont{Truscott}},
  \bibinfo{author}{\bibfnamefont{K.}~\bibnamefont{Baldwin}},
  \bibinfo{author}{\bibfnamefont{G.}~\bibnamefont{Birkl}},
  \bibinfo{author}{\bibfnamefont{P.}~\bibnamefont{Cancio}}, \bibnamefont{and}
  \bibinfo{author}{\bibfnamefont{M.}~\bibnamefont{Trippenbach}},
  \bibinfo{journal}{Rev. Mod. Phys.} \textbf{\bibinfo{volume}{84}},
  \bibinfo{pages}{175} (\bibinfo{year}{2012}).

\bibitem[{\citenamefont{Blaum et~al.}(2008)\citenamefont{Blaum, Geithner,
  Lassen, Lievens, Marinova, and Neugart}}]{2008-Argon-Blaum-Isotope-Shift}
\bibinfo{author}{\bibfnamefont{K.}~\bibnamefont{Blaum}},
  \bibinfo{author}{\bibfnamefont{W.}~\bibnamefont{Geithner}},
  \bibinfo{author}{\bibfnamefont{J.}~\bibnamefont{Lassen}},
  \bibinfo{author}{\bibfnamefont{P.}~\bibnamefont{Lievens}},
  \bibinfo{author}{\bibfnamefont{K.}~\bibnamefont{Marinova}}, \bibnamefont{and}
  \bibinfo{author}{\bibfnamefont{R.}~\bibnamefont{Neugart}},
  \bibinfo{journal}{Nuclear Physics A} \textbf{\bibinfo{volume}{799}},
  \bibinfo{pages}{30 } (\bibinfo{year}{2008}), ISSN \bibinfo{issn}{0375-9474}.

\bibitem[{\citenamefont{Cannon and Janik}(1990)}]{1990-Krypton-Isotope-Shofts}
\bibinfo{author}{\bibfnamefont{B.~D.} \bibnamefont{Cannon}} \bibnamefont{and}
  \bibinfo{author}{\bibfnamefont{G.~R.} \bibnamefont{Janik}},
  \bibinfo{journal}{Phys. Rev. A} \textbf{\bibinfo{volume}{42}},
  \bibinfo{pages}{397} (\bibinfo{year}{1990}).

\bibitem[{\citenamefont{Feldker et~al.}(2011)\citenamefont{Feldker, Schutz,
  John, and Birkl}}]{2011-isotope-shift}
\bibinfo{author}{\bibfnamefont{T.}~\bibnamefont{Feldker}},
  \bibinfo{author}{\bibfnamefont{J.}~\bibnamefont{Schutz}},
  \bibinfo{author}{\bibfnamefont{H.}~\bibnamefont{John}}, \bibnamefont{and}
  \bibinfo{author}{\bibfnamefont{G.}~\bibnamefont{Birkl}},
  \bibinfo{journal}{The European Physical Journal D}
  \textbf{\bibinfo{volume}{65}}, \bibinfo{pages}{257} (\bibinfo{year}{2011}),
  ISSN \bibinfo{issn}{1434-6060}.

\bibitem[{\citenamefont{Walhout et~al.}(1993)\citenamefont{Walhout, Megens,
  Witte, and Rolston}}]{1993-Isoshift-Xenon}
\bibinfo{author}{\bibfnamefont{M.}~\bibnamefont{Walhout}},
  \bibinfo{author}{\bibfnamefont{H.~J.~L.} \bibnamefont{Megens}},
  \bibinfo{author}{\bibfnamefont{A.}~\bibnamefont{Witte}}, \bibnamefont{and}
  \bibinfo{author}{\bibfnamefont{S.~L.} \bibnamefont{Rolston}},
  \bibinfo{journal}{Phys. Rev. A} \textbf{\bibinfo{volume}{48}},
  \bibinfo{pages}{R879} (\bibinfo{year}{1993}).

\bibitem[{\citenamefont{Zhao et~al.}(1991)\citenamefont{Zhao, Lawall, and
  Pipkin}}]{1991-he34-IsoShift}
\bibinfo{author}{\bibfnamefont{P.}~\bibnamefont{Zhao}},
  \bibinfo{author}{\bibfnamefont{J.~R.} \bibnamefont{Lawall}},
  \bibnamefont{and} \bibinfo{author}{\bibfnamefont{F.~M.}
  \bibnamefont{Pipkin}}, \bibinfo{journal}{Phys. Rev. Lett.}
  \textbf{\bibinfo{volume}{66}}, \bibinfo{pages}{592} (\bibinfo{year}{1991}).

\bibitem[{\citenamefont{Lu et~al.}(2013)\citenamefont{Lu, Mueller, Drake,
  N\"ortersh\"auser, Pieper, and Yan}}]{2013-laserProbing-nuclei-Lu}
\bibinfo{author}{\bibfnamefont{Z.-T.} \bibnamefont{Lu}},
  \bibinfo{author}{\bibfnamefont{P.}~\bibnamefont{Mueller}},
  \bibinfo{author}{\bibfnamefont{G.~W.~F.} \bibnamefont{Drake}},
  \bibinfo{author}{\bibfnamefont{W.}~\bibnamefont{N\"ortersh\"auser}},
  \bibinfo{author}{\bibfnamefont{S.~C.} \bibnamefont{Pieper}},
  \bibnamefont{and} \bibinfo{author}{\bibfnamefont{Z.-C.} \bibnamefont{Yan}},
  \bibinfo{journal}{Rev. Mod. Phys.} \textbf{\bibinfo{volume}{85}},
  \bibinfo{pages}{1383} (\bibinfo{year}{2013}).

\bibitem[{\citenamefont{Chen et~al.}(1999)\citenamefont{Chen, Li, Bailey,
  O'Connor, Young, and Lu}}]{1999-ATTA-Krypton}
\bibinfo{author}{\bibfnamefont{C.~Y.} \bibnamefont{Chen}},
  \bibinfo{author}{\bibfnamefont{Y.~M.} \bibnamefont{Li}},
  \bibinfo{author}{\bibfnamefont{K.}~\bibnamefont{Bailey}},
  \bibinfo{author}{\bibfnamefont{T.~P.} \bibnamefont{O'Connor}},
  \bibinfo{author}{\bibfnamefont{L.}~\bibnamefont{Young}}, \bibnamefont{and}
  \bibinfo{author}{\bibfnamefont{Z.-T.} \bibnamefont{Lu}},
  \bibinfo{journal}{Science} \textbf{\bibinfo{volume}{286}},
  \bibinfo{pages}{1139} (\bibinfo{year}{1999}).

\bibitem[{\citenamefont{Reichel et~al.}(2013)\citenamefont{Reichel, Kersting,
  Ritterbusch, Ebser, Bender, Purtschert, Oberthaler, and
  Aeschbach-Hertig}}]{2013-ArgonATTA}
\bibinfo{author}{\bibfnamefont{T.}~\bibnamefont{Reichel}},
  \bibinfo{author}{\bibfnamefont{A.}~\bibnamefont{Kersting}},
  \bibinfo{author}{\bibfnamefont{F.}~\bibnamefont{Ritterbusch}},
  \bibinfo{author}{\bibfnamefont{S.}~\bibnamefont{Ebser}},
  \bibinfo{author}{\bibfnamefont{K.}~\bibnamefont{Bender}},
  \bibinfo{author}{\bibfnamefont{R.}~\bibnamefont{Purtschert}},
  \bibinfo{author}{\bibfnamefont{M.}~\bibnamefont{Oberthaler}},
  \bibnamefont{and}
  \bibinfo{author}{\bibfnamefont{W.}~\bibnamefont{Aeschbach-Hertig}}, in
  \emph{\bibinfo{booktitle}{EGU General Assembly Conference Abstracts}}
  (\bibinfo{year}{2013}), vol.~\bibinfo{volume}{15}, p. \bibinfo{pages}{10901}.

\bibitem[{\citenamefont{Behr and Gorelov}(2014)}]{2014-Behr-beta}
\bibinfo{author}{\bibfnamefont{J.~A.} \bibnamefont{Behr}} \bibnamefont{and}
  \bibinfo{author}{\bibfnamefont{A.}~\bibnamefont{Gorelov}},
  \bibinfo{journal}{Journal of Physics G: Nuclear and Particle Physics}
  \textbf{\bibinfo{volume}{41}}, \bibinfo{pages}{114005}
  (\bibinfo{year}{2014}).

\bibitem[{\citenamefont{Wahlin}(2011)}]{2011-RFIS-Guide}
\bibinfo{author}{\bibfnamefont{E.}~\bibnamefont{Wahlin}},
  \emph{\bibinfo{title}{Model RFIS-100 RF ion source User's Manual}},
  \bibinfo{organization}{Beam Imaging Solutions} (\bibinfo{year}{2011}).

\bibitem[{\citenamefont{van~der Velden et~al.}(2004)\citenamefont{van~der
  Velden, Batelaan, te~Sligte, Beijerinck, and Vredenbregt}}]{Magnetron-2004}
\bibinfo{author}{\bibfnamefont{M.~H.~L.} \bibnamefont{van~der Velden}},
  \bibinfo{author}{\bibfnamefont{H.}~\bibnamefont{Batelaan}},
  \bibinfo{author}{\bibfnamefont{E.}~\bibnamefont{te~Sligte}},
  \bibinfo{author}{\bibfnamefont{H.~C.~W.} \bibnamefont{Beijerinck}},
  \bibnamefont{and} \bibinfo{author}{\bibfnamefont{E.~J.~D.}
  \bibnamefont{Vredenbregt}}, \bibinfo{journal}{Review of Scientific
  Instruments} \textbf{\bibinfo{volume}{75}} (\bibinfo{year}{2004}).

\bibitem[{\citenamefont{Daerr et~al.}(2011)\citenamefont{Daerr, Kohler,
  Sahling, Tippenhauer, Arabi-Hashemi, Becker, Sengstock, and
  Kalinowski}}]{2011-VUV-Krypton}
\bibinfo{author}{\bibfnamefont{H.}~\bibnamefont{Daerr}},
  \bibinfo{author}{\bibfnamefont{M.}~\bibnamefont{Kohler}},
  \bibinfo{author}{\bibfnamefont{P.}~\bibnamefont{Sahling}},
  \bibinfo{author}{\bibfnamefont{S.}~\bibnamefont{Tippenhauer}},
  \bibinfo{author}{\bibfnamefont{A.}~\bibnamefont{Arabi-Hashemi}},
  \bibinfo{author}{\bibfnamefont{C.}~\bibnamefont{Becker}},
  \bibinfo{author}{\bibfnamefont{K.}~\bibnamefont{Sengstock}},
  \bibnamefont{and} \bibinfo{author}{\bibfnamefont{M.~B.}
  \bibnamefont{Kalinowski}}, \bibinfo{journal}{Review of Scientific
  Instruments} \textbf{\bibinfo{volume}{82}}, \bibinfo{eid}{073106}
  (\bibinfo{year}{2011}).

\bibitem[{\citenamefont{Chen et~al.}(2001)\citenamefont{Chen, Bailey, Li,
  O'Connor, Lu, Du, Young, and Winkler}}]{2001-RF-Discharge-Krypton-Chen}
\bibinfo{author}{\bibfnamefont{C.~Y.} \bibnamefont{Chen}},
  \bibinfo{author}{\bibfnamefont{K.}~\bibnamefont{Bailey}},
  \bibinfo{author}{\bibfnamefont{Y.~M.} \bibnamefont{Li}},
  \bibinfo{author}{\bibfnamefont{T.~P.} \bibnamefont{O'Connor}},
  \bibinfo{author}{\bibfnamefont{Z.-T.} \bibnamefont{Lu}},
  \bibinfo{author}{\bibfnamefont{X.}~\bibnamefont{Du}},
  \bibinfo{author}{\bibfnamefont{L.}~\bibnamefont{Young}}, \bibnamefont{and}
  \bibinfo{author}{\bibfnamefont{G.}~\bibnamefont{Winkler}},
  \bibinfo{journal}{Review of Scientific Instruments}
  \textbf{\bibinfo{volume}{72}}, \bibinfo{pages}{271} (\bibinfo{year}{2001}).

\bibitem[{\citenamefont{Conrads and Schmidt}(2000)}]{2000ICP-CCP}
\bibinfo{author}{\bibfnamefont{H.}~\bibnamefont{Conrads}} \bibnamefont{and}
  \bibinfo{author}{\bibfnamefont{M.}~\bibnamefont{Schmidt}},
  \bibinfo{journal}{Plasma Sources Science and Technology}
  \textbf{\bibinfo{volume}{9}}, \bibinfo{pages}{441} (\bibinfo{year}{2000}).

\bibitem[{\citenamefont{Ritterbusch}(2009)}]{2009-Florian}
\bibinfo{author}{\bibfnamefont{F.}~\bibnamefont{Ritterbusch}}, Ph.D. thesis,
  \bibinfo{school}{Kirchhoff Institute for Physics} (\bibinfo{year}{2009}).

\bibitem[{\citenamefont{Welte}(2011)}]{2011-Welte}
\bibinfo{author}{\bibfnamefont{J.}~\bibnamefont{Welte}}, Ph.D. thesis,
  \bibinfo{school}{Ruperto-Carola-University of Heidelberg}
  (\bibinfo{year}{2011}).

\bibitem[{\citenamefont{Sulai}(2011)}]{2011-Suil}
\bibinfo{author}{\bibfnamefont{I.~A.} \bibnamefont{Sulai}}, Ph.D. thesis,
  \bibinfo{school}{University of chicago} (\bibinfo{year}{2011}).

\bibitem[{\citenamefont{Phillips et~al.}(1985)\citenamefont{Phillips, Prodan,
  and Metcalf}}]{1985-Metcalf-Book}
\bibinfo{author}{\bibfnamefont{W.~D.} \bibnamefont{Phillips}},
  \bibinfo{author}{\bibfnamefont{J.~V.} \bibnamefont{Prodan}},
  \bibnamefont{and} \bibinfo{author}{\bibfnamefont{H.~J.}
  \bibnamefont{Metcalf}}, \bibinfo{journal}{J. Opt. Soc. Am. B}
  \textbf{\bibinfo{volume}{2}}, \bibinfo{pages}{1751} (\bibinfo{year}{1985}).

\bibitem[{\citenamefont{Ralchenko et~al.}(2008)\citenamefont{Ralchenko,
  Kramida, Reader, and Team}}]{2008-NIST}
\bibinfo{author}{\bibfnamefont{Y.}~\bibnamefont{Ralchenko}},
  \bibinfo{author}{\bibfnamefont{A.~E.} \bibnamefont{Kramida}},
  \bibinfo{author}{\bibfnamefont{J.}~\bibnamefont{Reader}}, \bibnamefont{and}
  \bibinfo{author}{\bibfnamefont{N.~A. S.~D.} \bibnamefont{Team}},
  \emph{\bibinfo{title}{{NIST Atomic Spectra Database (version 3.1.5)}}}
  (\bibinfo{year}{2008}).

\bibitem[{\citenamefont{A.Kramida et~al.}()\citenamefont{A.Kramida,
  {Yu.Ralchenko}, J.Reader, and {and NIST ASD Team}}}]{2014-NIST_ASD}
\bibinfo{author}{\bibnamefont{A.Kramida}},
  \bibinfo{author}{\bibnamefont{{Yu.Ralchenko}}},
  \bibinfo{author}{\bibnamefont{J.Reader}}, \bibnamefont{and}
  \bibinfo{author}{\bibnamefont{{and NIST ASD Team}}}.

\bibitem[{\citenamefont{Pauly}(2000)}]{pauly2000atom}
\bibinfo{author}{\bibfnamefont{H.}~\bibnamefont{Pauly}},
  \emph{\bibinfo{title}{Atom, Molecule, and Cluster Beams I: Basic Theory,
  Production and Detection of Thermal Energy Beams}}
  (\bibinfo{publisher}{Springer}, \bibinfo{year}{2000}).

\bibitem[{\citenamefont{Lubotzky}(2013)}]{2013-Boaz}
\bibinfo{author}{\bibfnamefont{B.}~\bibnamefont{Lubotzky}}, Master's thesis,
  \bibinfo{school}{Hebrew University of Jerusalem} (\bibinfo{year}{2013}).

\bibitem[{\citenamefont{Welte et~al.}(2010)\citenamefont{Welte, Ritterbusch,
  Steinke, Henrich, Aeschbach-Hertig, and
  Oberthaler}}]{2010-Welte-florian-Paper}
\bibinfo{author}{\bibfnamefont{J.}~\bibnamefont{Welte}},
  \bibinfo{author}{\bibfnamefont{F.}~\bibnamefont{Ritterbusch}},
  \bibinfo{author}{\bibfnamefont{I.}~\bibnamefont{Steinke}},
  \bibinfo{author}{\bibfnamefont{M.}~\bibnamefont{Henrich}},
  \bibinfo{author}{\bibfnamefont{W.}~\bibnamefont{Aeschbach-Hertig}},
  \bibnamefont{and} \bibinfo{author}{\bibfnamefont{M.~K.}
  \bibnamefont{Oberthaler}}, \bibinfo{journal}{New Journal of Physics}
  \textbf{\bibinfo{volume}{12}}, \bibinfo{pages}{065031}
  (\bibinfo{year}{2010}).

\bibitem[{\citenamefont{Hotop}(1996)}]{1996-hotop}
\bibinfo{author}{\bibfnamefont{H.}~\bibnamefont{Hotop}}, in
  \emph{\bibinfo{booktitle}{Atomic, Molecular, and Optical Physics: Atoms and
  Molecules}}, edited by
  \bibinfo{editor}{\bibfnamefont{F.}~\bibnamefont{Dunning}} \bibnamefont{and}
  \bibinfo{editor}{\bibfnamefont{R.~G.} \bibnamefont{Hulet}}
  (\bibinfo{publisher}{Academic Press}, \bibinfo{year}{1996}), vol.
  \bibinfo{volume}{29, Part B}, pp. \bibinfo{pages}{191 -- 215}.

\bibitem[{\citenamefont{Cobas and Lamb}(1944)}]{1944-Metastable-Metal}
\bibinfo{author}{\bibfnamefont{A.}~\bibnamefont{Cobas}} \bibnamefont{and}
  \bibinfo{author}{\bibfnamefont{W.~E.} \bibnamefont{Lamb}},
  \bibinfo{journal}{Phys. Rev.} \textbf{\bibinfo{volume}{65}},
  \bibinfo{pages}{327} (\bibinfo{year}{1944}).

\bibitem[{\citenamefont{Dunning et~al.}(1975)\citenamefont{Dunning, Rundel, and
  Stebbings}}]{1975-dunning}
\bibinfo{author}{\bibfnamefont{F.~B.} \bibnamefont{Dunning}},
  \bibinfo{author}{\bibfnamefont{R.~D.} \bibnamefont{Rundel}},
  \bibnamefont{and} \bibinfo{author}{\bibfnamefont{R.~F.}
  \bibnamefont{Stebbings}}, \bibinfo{journal}{Review of Scientific Instruments}
  \textbf{\bibinfo{volume}{46}} (\bibinfo{year}{1975}).

\bibitem[{\citenamefont{Ashmore}(2005)}]{2005-Ashmore}
\bibinfo{author}{\bibfnamefont{J.~P.} \bibnamefont{Ashmore}}, Ph.D. thesis,
  \bibinfo{school}{{Griffith University}} (\bibinfo{year}{2005}).

\bibitem[{\citenamefont{Brand et~al.}(1992)\citenamefont{Brand, Furst, Gay, and
  Schearer}}]{1992-Missoury-Sat-Quench}
\bibinfo{author}{\bibfnamefont{J.~A.} \bibnamefont{Brand}},
  \bibinfo{author}{\bibfnamefont{J.~E.} \bibnamefont{Furst}},
  \bibinfo{author}{\bibfnamefont{T.~J.} \bibnamefont{Gay}}, \bibnamefont{and}
  \bibinfo{author}{\bibfnamefont{L.~D.} \bibnamefont{Schearer}},
  \bibinfo{journal}{Review of Scientific Instruments}
  \textbf{\bibinfo{volume}{63}}, \bibinfo{pages}{163} (\bibinfo{year}{1992}).

\bibitem[{\citenamefont{Allred et~al.}(2010)\citenamefont{Allred, Reeves,
  Corder, and Metcalf}}]{2010-helium-lithography}
\bibinfo{author}{\bibfnamefont{C.~S.} \bibnamefont{Allred}},
  \bibinfo{author}{\bibfnamefont{J.}~\bibnamefont{Reeves}},
  \bibinfo{author}{\bibfnamefont{C.}~\bibnamefont{Corder}}, \bibnamefont{and}
  \bibinfo{author}{\bibfnamefont{H.}~\bibnamefont{Metcalf}},
  \bibinfo{journal}{Journal of Applied Physics} \textbf{\bibinfo{volume}{107}},
  \bibinfo{eid}{033116} (\bibinfo{year}{2010}).

\end{thebibliography}

\end{document}